**Interplay between disorder driven Non-Fermi-liquid behavior and magnetism in $Ce_{0.24}La_{0.76}Ge$ compound**

Karan Singh and K. Mukherjee*

School of Basic Sciences, Indian Institute of Technology Mandi, Mandi 175005, Himachal Pradesh, India

Email: kaustav@iitmandi.ac.in

**Abstract -** In this work, we investigate the magnetic, heat capacity and electrical transport properties of $Ce_{0.6}La_{0.4}Ge$ and $Ce_{0.24}La_{0.76}Ge$ compounds. Our results show that two antiferromagnetic transitions (~ at 4.7 and 2.7 K) exhibited by $Ce_{0.6}La_{0.4}Ge$ are suppressed below 1.8 K for $Ce_{0.24}La_{0.76}Ge$. Interestingly, for $Ce_{0.24}La_{0.76}Ge$, susceptibility, heat capacity and electrical resistivity vary with temperature as: $T^{0.75}$, $T^{0.5}$ and $T^{1.6}$ respectively. The observation of such anomalous temperature variation suggests to the Non-Fermi-liquid (NFL) behavior due to the presence of disordered 4$f$ spins due to Ce-site dilution. Under the application of magnetic field, it is noted that a crossover from the NFL to a magnetic state occurs around 2 Tesla, where, short-range correlations among the spins is prevalent due to the dominance of coupling between the magnetic moments via conduction electrons. Magnetoresistance scaling indicates that behavior of disorder driven NFL state is described by the dynamical mean field theory of the spin glass quantum critical point.

**Introduction: -** In the area of research on 4$f$ electron based compounds, a challenge which remains elusive is to understand the physical properties of materials where singular interaction is mediated by soft collective modes. This investigation is important as the obtained results violate the applicability of Fermi-liquid (FL) theory [1-2]. The FL theory forms a basis to understand the electronic properties and according to this theory, electrical resistivity ($\rho$) varies as the square of temperature ($T$) and the ratio of heat capacity ($C$) and temperature i.e. $C/T$ is constant [3-4]. However, it has been noted in many 4$f$ electron based compounds that the FL theory breaks down and shows a Non-Fermi-Liquid (NFL) behavior near quantum critical point (QCP). The $\rho$ of such compound shows a linear behavior with $T$ while, $T$ variation of $C/T$ is logarithmic [5-8]. The observation of such NFL behavior might arise due to the presence of soft order parameter fluctuations resulting in singular interactions mediated between electrons [9-11]. In some compounds exhibiting NFL behavior, ρ is reported to vary as $T^{0.5}$ [12-14]. The observation of such features is explained on the basis of multi-channel Kondo model (such as two-channel Kondo model), which is expected in 4$f$ electron based compounds [15-18]. The conduction electron is further coupled to the 4$f$ electron state via the spin-orbit coupling which results in stretching and squashing of conduction electron orbital. This effect might be responsible for the splitting of six-fold degeneracy of $J$=5/2 ground multiplets into the quartet excited state along with the doublet ground state under crystalline electric field [17-21]. The coupling of the quartet of conduction electron with localized moments can be described by the two-channel Kondo model which a result in ρ varies with $T^{0.5}$ [22]. However, in disorder driven NFL compound, ρ varies with $T^{1.5}$ [23-25]. For understanding this behavior of resistivity, recently, spin glass (SG) model in the frame of the dynamical field theory has been predicted [24, 26]. This theory explains such variation in resistivity on the basis of the presence of disorder when exchange energy is comparable to Kondo energy. But, when the exchange energy is not equal to the Kondo energy, presence of disorder might result in the formation of the SG phase.

CeGe is an interesting 4$f$ electron based system and it undergoes two orderings at $T^I$ (~ 10.7 K) and $T^{II}$ (~ 4.6 K) at low magnetic field. The former one is due to antiferromagnetic ordering while the later one arises due to some spin rearrangement [27]. The high magnetic field dependent studies of physical properties on CeGe revealed to the absence of signature of FL or NFL state [28-31], which is generally observed for other Ce-based compounds [7]. Hence, it would be of interest to investigate the physical properties of CeGe by dilution of Ce-site by a nonmagnetic ion, as there are no such literature reports about these investigations on this compound; to the best of our knowledge. In this letter, we focus on magnetization, thermodynamics and transport properties of $Ce_{0.6}La_{0.4}Ge$ and $Ce_{0.24}La_{0.76}Ge$ compounds in low temperature region. We choose $Ce_{0.24}La_{0.76}Ge$ because the ordering temperature of the compound CeGe decrease with the increasing La-substitution and it is expected that the magnetic transition temperature may be suppressed below 1.8 K in the $Ce_{0.24}La_{0.76}Ge$. Also, the Kondo energy scale may be approximately comparable to the magnetic exchange energy in $Ce_{0.24}La_{0.76}Ge$. Moreover, introduction of La will induce disorder and will also tune the hybridization of localized moments with conduction electrons. This may result in the observation of exotic physical properties. Our results revealed that the magnetic ordering temperature of $Ce_{0.6}La_{0.4}Ge$ is shifted down in temperature while for $Ce_{0.24}La_{0.76}Ge$ it is suppressed below 1.8 K. Interestingly, for Ce$_{0.24}$La$_{0.76}$Ge, the variation of heat capacity and resistivity shows the presence of NFL behavior which arises as a result of interplay of disorder effect due to 4$f$ spins (due to site dilution effect) along with balanced Kondo and magnetic exchange energy. Above 2 Tesla, our results divulge the presence of field induced magnetic state arising due to the presence of short range correlations among the magnetic moments. Magnetoresistance scaling indicates that behavior of the disorder driven NFL state is described by the dynamical mean field theory of the spin glass quantum critical point.

**Experimental-** High quality polycrystalline compounds Ce$_{0.6}$La$_{0.4}$Ge (La_0.4) and Ce$_{0.24}$La$_{0.76}$Ge (La_0.76) are prepared by arc melting stoichiometric amounts of respective high purity elements (>99.9%) in an atmosphere of argon. The compounds are re-melted several times to ensure homogeneity and weight losses after final melting is insignificant (<0.5%). X-ray diffraction studies at room temperature established the single phase nature of these compounds. The indexed XRD is shown in Fig. 1 (a). The compounds crystallize in orthorhombic structure. For Ce$_{0.6}$La$_{0.4}$Ge, the obtained lattice parameters are $a$ = 8.398(4) Å, $b$ = 4.098(6) Å, $c$ = 6.059(7) Å and $V$ = 208.585 Å$^3$, whereas, for Ce$_{0.24}$La$_{0.76}$Ge the parameters are $a$ = 8.434(4) Å, $b$ = 4.116(9) Å, $c$ = 6.089(7) Å and $V$ = 211.454 Å$^3$. From the lattice parameters it is noted that with an increase in La-substitution, the unit cell volume increases. The peak shifts towards lower angle side (inset of Fig. 1(a)), thereby establishing the dopant goes to the Ce-site. We have also performed the energy dispersive spectroscopy (EDAX) to get confirmation about the average atomic stoichiometry, which is found to be in accordance with the expected values. Temperature and magnetic field dependent heat capacity and resistivity are performed using Physical Property Measurement System (PPMS), while temperature dependent magnetization studies are performed using Magnetic Property Measurement System (MPMS); both from Quantum design, USA.

**Results and discussions -** Fig. 1 (b) shows the temperature dependent DC susceptibility for La_0.4 and La_0.76 compounds at 0.005 Tesla under zero field cooling (ZFC) condition. The parent compound CeGe undergoes two antiferromagnetic transitions approximately at 10.7 and 7.6 K [27]. These transitions are shifted to 4.7 ($T^I$) and 2.7 K ($T^{II}$) for La_0.4 (shown inset of Fig. 1 (b)). For the extreme composition La_0.76, the transition temperature is suppressed below 1.8

K. Temperature response of the AC susceptibility in the presence of a superimposed DC field of 0.005 Tesla for both the compound is shown in Fig. 1 (c). From the figure, it is noted that the real part of AC susceptibility ($\chi'_{ac}$) for La_0.4 compound shows a sharp peak around 4.7 K followed by a weak anomaly around 2.7 K (shown by an arrow) which is in accordance with the transition temperature noted from DC susceptibility. For La_0.76 compound, no such peaks are observed. For both compounds, imaginary part of AC susceptibility ($\chi''_{ac}$) is insignificant and is not detected by the instrument. Fig. 1 (d) shows the magnetic field dependent isothermal magnetization at lowest attainable temperature of 1.8 K for both compounds. The magnetization increases with increasing magnetic field; however at low field, it is observed that slope of curve changes with La-doping resulting in increment of magnetic moment. Also, it is observed that magnetic hysteresis is absent in both of these compounds (not shown).

Fig. 2 (a) shows the temperature dependent heat capacity ($C$) for La_0.4 and La_0.76 compounds at 0 Tesla. For La_0.4 compound, a peak is observed around $T^I$ followed by a weak anomaly around $T^{II}$ (shown by an arrow). The curve is featureless for La_0.76 compound because the transition temperature is shifted below 1.8 K. For both compounds, it is noted that with decreasing temperature $C$ increases and attains broad maxima around 155 K. Such feature possibly arises due to the presence of incoherent Kondo scattering along with the CEF effect [32]. Upper inset of Fig. 2 (a) shows the heat capacity divided by temperature as a function of $T^2$. To subtract the phonon contribution, we fit the equation: $C/T = \gamma + \beta T^2$ in the temperature range of 10 - 30 K. In this equation, $\gamma$ and $\beta$ are the electronic and phonon contributions of the heat capacity. The obtained values of $\gamma$ are 94.2 and 2.4 mJ/mol-K$^2$ for La_0.4 and La_0.76 compounds respectively. It is noted that $\gamma$ decreases with increasing La-substitution resulting in decrease of screening effect of magnetic moments. This effect is due to the fact that the increase in unit cell volume with increasing La-substitution results in the decrement of $Jn(E_f)$ (where $J$ is exchange interaction parameter of f-d hybridization and $n(E_f)$ is electronic states per unit volume at Fermi level) [33]. Lower inset of Fig. 2 (a) shows the temperature dependent $C'/T$ for La_0.76 compound, which is obtained after subtracting the phonon contribution from heat capacity up to the lowest temperature using the equation: $C' = C - \beta T^3$. The curve increases with decreasing temperature and it is fitted (in temperature range 1.8 – 3.5 K) using the equation: $C'/T = \gamma_0 - \alpha T^b$, where $b$ is temperature exponent, $\gamma_0$ and $\alpha$ are the free fitting parameters [7]. Best fitting is obtained for $b = 0.5 \pm 0.03$, $\gamma_0$ and $\alpha$ are ~ $0.890 \pm 0.01$ and $0.45 \pm 0.01$ respectively. The value of *b* indicates to the presence of NFL behavior in the low temperature regime. This feature might arise due to critical spin fluctuation because of equivalent strength of Kondo and magnetic exchange energy [25]. Fig. 2 (b) shows the temperature dependent normalized resistivity ($\rho/\rho_{200}$) for both compounds at 0 Tesla. With decreasing temperature, resistivity decreases which is a characteristic of metallic behavior. However, for La_0.4 compound, a non-linear deviation is noted below 160 K. Such feature can possibly arise due to the effect of partial screening of magnetic moments via conduction electrons and/or CEF. This non-linear deviation is reduced for La_0.76 due to the decrease of screening of magnetic moments which is also in analogy with the analysis of heat capacity data. The decrease of screening effect of magnetic moments suggests that degeneracy of the 4*f* level might play an important role in the electronic and thermodynamic properties in low temperature region [32, 34]. For La_0.4 compound, it is noted that resistivity rises around $T^I$, followed by a maximum around $T^{II}$ (upper inset of Fig.2 (b)). Similar effect has been reported for CeGe which undergo gap opening at $T^I$ and some spin rearrangement at $T^{II}$ [27]. Below $T^{II}$, coherence due to some spin rearrangement may arise, which results in a decrease in resistivity. For La_0.76 compound, no such upturn is noted. Hence, it can be said that with La-substitution gap opening is shifted to lower temperature (below 1.8 K) or is suppressed. For La_0.76 compound, variation of resistivity in the temperature range of 1.8 to 3.5 K is fitted with the following equation: $\rho = \rho_0 + AT^n$, where $\rho_0$ is the residual resistivity and *A* is associated with the scattering of conduction electrons and *n* is the temperature exponent [24]. Best fitting is obtained for $n = 1.6 \pm 0.02$. Lower inset of Fig. 2 (b) shows ρ plotted as a function of $T^{1.6}$ which illustrates a linear behavior (arrow indicates to deviation in curve around 3.5 K). The value of the exponent *n* also indicates to the observation of NFL behavior in this compound. As per literature reports, ρ varies with *T* due to critical spin fluctuations [5-6]. However, the observation of unusual variation of ρ (varies with $T^{1.6}$) suggests that the observed NFL behavior could be caused by the interplay of disorder effect due to 4*f* spins (due to site dilution effect) along with critical spin fluctuations. Similar effect has been reported for other compounds like $CeNi_2Ge_2$ and $CeCu_2Si_2$ [35]. Hence, it can be said that with increasing La-substitution the hybridization of magnetic moments with the conduction electrons is modified resulting in observation of unusual NFL behavior in La_0.76 compound. In order to further explore this behavior, we have studied the temperature response of heat capacity and resistivity under the influence of magnetic field up to 14 Tesla.

Fig. 3 (a) shows the temperature dependent $C'/T$ under different magnetic fields. As the magnetic field increases up to 2 Tesla, the nature of slope of $C'/T$ remains unchanged, indicating the persistence of NFL behavior. As the magnetic field is increased further, a curvature is noted around 2.2 K for 3 Tesla which indicates towards a new magnetic state due to the dominance of short range correlations (SRC) among the localized magnetic moments which results in suppression of NFL behavior. Above 3 Tesla, this curvature becomes more prominent with increasing magnetic field. Weak signature of the curvature is also noted in the temperature response of $M/H$ curves at high fields (inset of Fig. 3 (a)). Temperature dependent entropy ($S'$) curve at 0 Tesla is shown in the inset of Fig.3 (b). The $S'$ decreases with decreasing temperature and obtained entropy is about 0.30 (0.052 Rln2) around 3.5 K. The observed entropy is very small as compared to the compound which shows the multi-channel Kondo effect [36]. Hence, it can be said that existence of adequate disorder by the Ce-site dilution might be responsible for such low entropy [36]. The observation of NFL behavior is further analyzed by the scaling of the $C'/T$ with the characteristic temperature $T^*$, which depends upon the strength of Kondo energy [11, 35]. Fig. 3 (b) shows the scaled $C'/(T/T^*)$ with the $T/T^*$ in field range 0-2 Tesla. This scaling gives further evidence about the presence of NFL behavior in this compound. Hence, it can be said that substitution of La at the Ce-site results in site dilution effect and at typical concentration i.e. at La_0.76, presence of sufficient disorder (as compared to CeGe) leads to the observed NFL behavior.

Fig. 4 (a) shows the temperature dependent resistivity $\rho$ under different fields. The resistivity decreases linearly with decreasing temperature for all fields due to metallic behavior of the compounds. In the low temperature regime, the curves are fitted with equation: $\rho = \rho_0 + AT^n$ (not shown). Inset of Fig. 4 (a) shows the magnetic field dependent residual resistivity ($\rho_0$). It is observed that $\rho_0$ increase up to 1 Tesla (approximately) due to increasing disorder. At high fields, $\rho_0$ decreases remarkably. It might be due to the dominance of SRC because of the presence of strong exchange coupling among the localized magnetic moments. It is also to be noted that the obtained value of the exponent $n$ is quite away from 2 (the value of $n$ for FL state) at high fields. Hence, it can be said that dominance of exchange energy impedes the recovery of FL state, which is also observed in other Ce-based compounds [24]. Possibly the development of SRC among the spins impede the recovery of FL state. To probe the exact nature of this magnetic state, low temperature neutron diffraction measurements are warranted which is beyond the scope of this manuscript. Interestingly, as per literature reports, disorder driven NFL state is analyzed by the dynamical mean field theory of spin glass quantum critical point (SG QCP) [24]. In the absence of disorder, the dynamical mean field theory describes the FL behavior in the low temperature region. However, the presence of significant disorder results in the inhomogeneous scattering which affects the electrical transport properties and observation of NFL behavior [23]. Hence, in our case, the presence of disorder along with coupling of conduction electron with critical spin fluctuation is described by SG QCP, which has been predicted to understand such anomalous variation of resistivity. Also, it has been reported that the physical properties depends on the effective distance to the QCP ($\Delta (T, H)$) and the resistivity in form of scaling: $\rho (T, H) - \rho (0, H) \propto t^{3/2} \Psi (t/\Delta)$, where $t = T/T_0$ ($T_0$ is the temperature which depends upon $J$) and $\Psi (x)$ is the scaling function [24]. For disorder driven NFL, the equation in the preceding line describes the $T^{3/2}$ dependence of resistivity. $\Delta$ is described as: $\Delta = \Delta_0 + 2(\Delta_0)^{1/2} t \{[1 + t/2(3)^{1/2}\Delta_0]^{1/2} - 1\}$, where $\Delta_0 = r + (H/H_0)^2$; $r (= 1-J/J_c)$ is the measure of the distance to the realistic QCP and $H_0 = J_c/(g\mu_B)$ ($g$ is the gyromagnetic ratio of the Ce ion, taken as 2 and $\mu_B$ is the Bohr magnetron). For the critical coupling (at the QCP), the exchange energy $J$ become $J_c$ (critical exchange energy) and $\Delta_0$ approximately equals to $(H/H_0)^2$. The susceptibility ($\chi = M/H$) for the critical coupling has the NFL behavior and is described as: $\chi=\chi(0)–cT^\beta$, for $T/T_0 \leq 1$, where $\chi(0)$ ($= (g\mu_B^2)/J_c$) is the zero temperature susceptibility and $c (= (g\mu_B^2)/J_c (T_0)^{0.75})$ is a coefficient [26]. A best fitting is obtained for $\beta = 0.75 \pm 0.01$. Inset of the Fig. 4 (b) show the linear fitting of $M/H$ with $T^{0.75}$ at 1 Tesla. The obtained value $J_c$ and $T_0$ are approximately 10.2 K and 7 K respectively. Also, from the equation: $\chi=\chi (0) – cT^\beta$, we estimate the characteristic field $H_0$ ($= \mu_B /\chi (0)$) which is near to 7.6 Tesla. Upper inset of Fig. 4 (b) shows the temperature dependent $\rho (T, H) - \rho (0, H)$ curves at magnetic fields $0 – 1$ Tesla. From these curves a best scaling of log $[\rho (T, H) - \rho (0, H)]/t^{1.5}$ vs. log $(t/\Delta)$ is obtained for the $T_0 = 7$ K, $H_0 = 9.6$ Tesla (we got a better scaling for 9.6 Tesla instead of 7.6 Tesla) and $r = 0.385$ (Fig. 4 (b)). Similar form of scaling has been reported in Ce(Ru$_{0.5}$Rh$_{0.5}$)$_2$Si$_2$, which also shows the disorder driven NFL behavior [24]. The scaling of the resistivity indicates that the behavior of this compound is described by the dynamical mean field theory of SG QCP. Also, we calculate the value of $\delta J = (J_c – J) = rJ_c = 3.9$ K. The $\delta J$ determined from scaling is close to the theoretical estimated value in Ref [26]. It has reported that for $J \leq J_c$, there are two possibilities with decreasing temperature: FL behavior or a magnetic state, depending on the factor $Jn (E_f)$ [33]. For La_0.76 compound, the FL behavior has been ruled out and our results indicate that field induced a new magnetic state which arises due to the presence of SRC among the magnetic moments. Similar type of behavior is also reported in Ref [37]. Above 2 Tesla, NFL behavior is suppressed and these weak interactions strengthen, leading to the growth of SRC among spin at high fields. This SRC among the spins results in the development of a new magnetic state at high fields. Thus, it can be said that in La_0.76 compound, the NFL phase is below 3.5 K and around 2 Tesla. In this phase $M/H$, $C'/T$ and ρ varies with $T^{0.75}$, $T^{0.5}$ and $T^{1.6}$ respectively. The behavior of NFL phase has been described by dynamical mean field theory of SG QCP. The phase above 2 Tesla and below the temperature where new anomaly is observed in the heat capacity curve for respective field is a new magnetic phase. In this phase, SRC among spins is prevalent due to dominance of interaction between the localized magnetic moments via conduction electrons because of the presence of a strong exchange coupling. The temperature of the anomaly increases with increasing magnetic field due to an increase of the coupling strength.

**Summary:** In this work, we have studied the low temperature magnetic, thermodynamic and transport properties ofLa_0.4 and La_0.76 compounds. It is observed that with increasing La-substitution, the ordering temperature is shifted down in temperature for La_0.4 (as compared to CeGe) and is suppressed below 1.8 K for La_0.76. Interestingly, temperature variation of resistivity and heat capacity suggest to the presence of unusual NFL behavior which arises due to the presence of disorder. As the magnetic field is increased above 2 Tesla, field induced new magnetic state is noted which arises due to the presence of the strong short range correlations among the magnetic moments. The magnetoresistance scaling indicates that the behavior of the disorder driven NFL state is described by the dynamical mean field theory of the spin glass quantum critical point.

[36] Yamane Y. *et al.*, *Phys. Rev. Lett.* **121** (2018) 077206

[37] Taniguchi T. *et al.*, *J. Phys. Soc. Jpn.* **68** (1999) 2026

**Figures:**

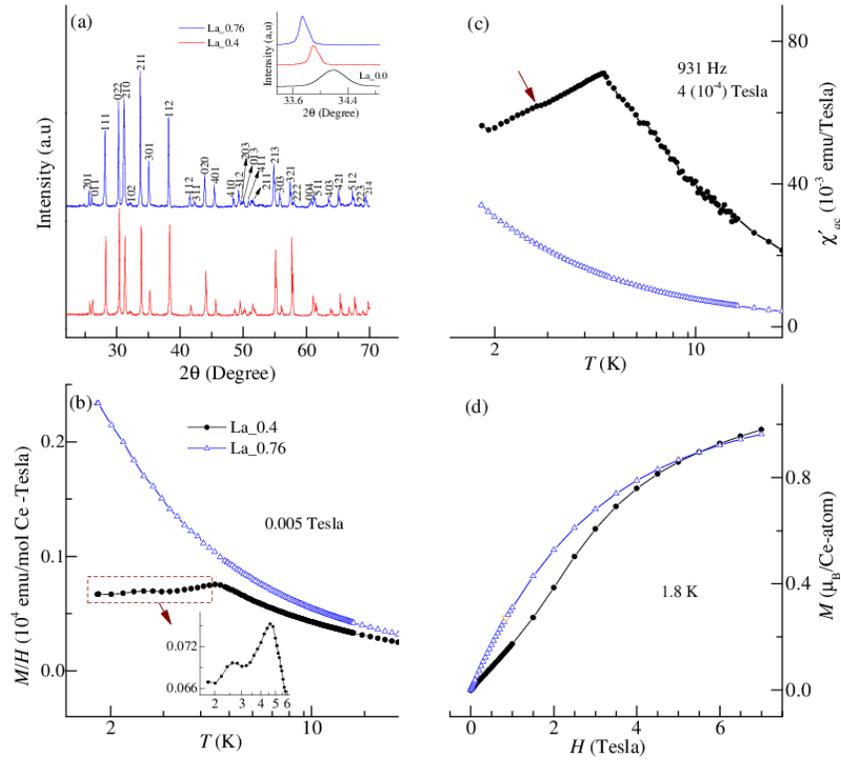

**Fig. 1:** (a) Indexed XRD patterns of La_0.4 and La_0.76 compounds. Inset: The region around the main peaks in an expanded form to show a gradual shift of diffraction lines with changing composition. The curve for La_0.0 (CeGe) compound is taken from Ref [27] for comparison. (b) Temperature ($T$) dependent DC susceptibility ($M/H$) at 0.005 Tesla under ZFC condition for La_0.4 and La_0.76 compounds. Inset: Magnified plot of the same figure in the temperature range 1.8- 6 K. (c) $T$ dependent real part of AC susceptibility ($\chi'_{ac}$) at superimposed DC field of 0.005 Tesla at 931 Hz in presence of $4(10^{-4})$ Tesla AC field. Arrow shows the weak anomaly in the $\chi_{ac}$ curve near 2.5 K. (d) Isothermal magnetization ($M$) plotted as a function of magnetic field at 1.8 K.

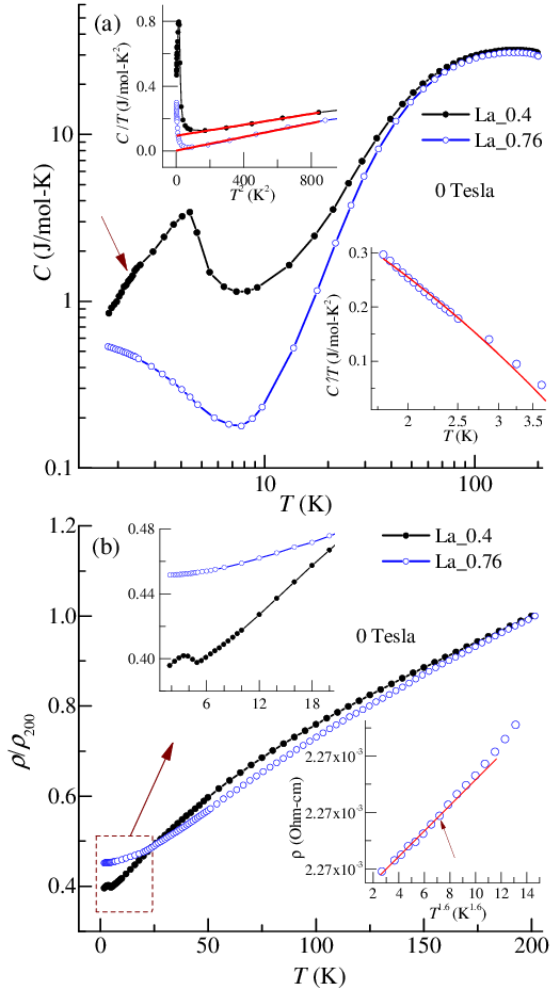

**Fig.2:** (a) $T$ dependent heat capacity ($C$) at 0 Tesla. Arrow shows the weak anomaly in $C$ curve near 3.5 K. Upper inset: $T^2$ dependent $C/T$ at 0 Tesla. Red line shows the linear fitting above ordering temperature, which is extrapolated to lowest temperature. Lower inset: $T$ dependent $C'/T$ at 0 Tesla. Red line shows the fitting using equation $C'/T = \gamma_0 - \alpha T^b$. (b) $T$ dependent normalized resistivity ($\rho/\rho_{200}$) at 0 Tesla. Upper inset: Magnified plot of the same figure in the temperature range 1.8-20 K. Lower inset: $T^{1.6}$ dependent $\rho$ at 0 Tesla. Red line shows the fitting using equation $\rho = \rho_0 + AT^n$. Arrow shows the deviation in $\rho$ near 3.5 K due to the onset of NFL region.

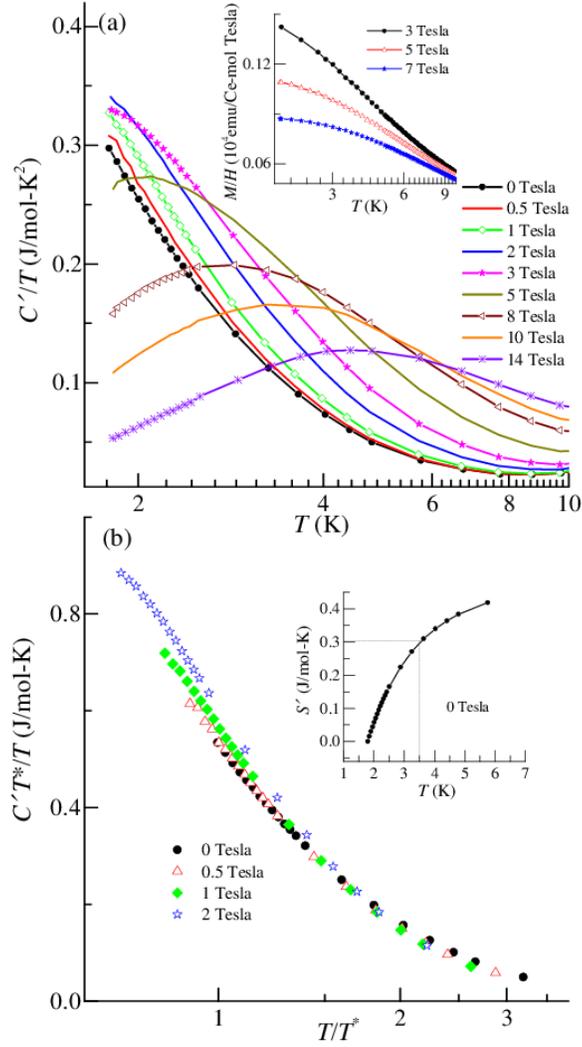

**Fig. 3:** (a) $T$ dependent heat capacity $C'/T$ in field range 0-14 Tesla for La_0.76 compound. Inset: $T$ (logarithmic scale) dependent DC susceptibility ($M/H$) at high fields. (b) $C'T^*/T$ versus $T/T^*$ (logarithmic scale) at different fields (0-2T). $T^*$ is the characteristic temperature and is taken as 1.8, 2, 2.2 and 2.6 K for $H = 0$, 0.5, 1 and 2 Tesla respectively. Inset: $T$ dependent entropy ($S'$) at 0 Tesla

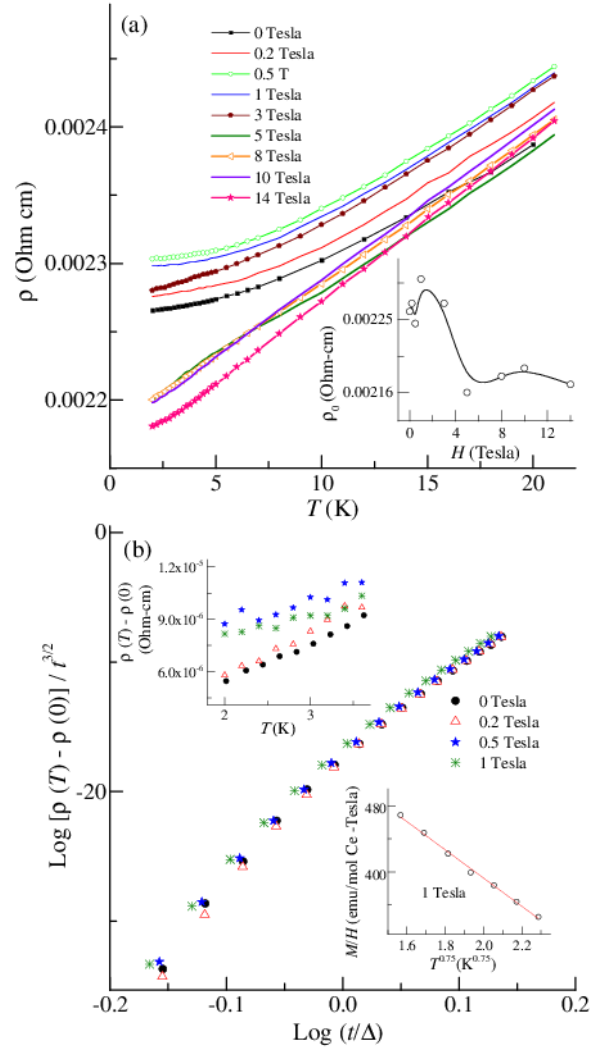

**Fig. 4:** (a) $T$ dependent resistivity in field range 0-14 Tesla for La_0.76 compound. Inset: Magnetic field dependent residual resistivity ($\rho_0$). (b) Scaling plots of the resistivity in low temperature region and field range 0-1 Tesla. Upper inset: $T$ dependent $\rho-\rho_0$ in field range 0-1 Tesla. Lower inset: $T$ dependent $M/H$ at 1 Tesla. Red line shows the fitting using the equation $\chi = \chi(0) - cT^\beta$.